# 3D Migration Aperture and Formula Connecting Dips of Prestack Time Migrated and Unmigrated Data

*Jagmeet Singh, Cewell, ONGC, India*

## Abstract

Formula connecting dips of unmigrated 2D and prestack time migrated constant offset gathers was recently proposed (Singh, 2024). A handy formula for 2D prestack migration aperture was also derived. We extend the same work to the 3D case, that is, we derive prestack time migration aperture, and formula connecting dips as a function of dip of a planar reflector and azimuth w.r.t. the dip direction.

## Introduction

Recently, Singh (2024) extended Chun and Jacewitz's formula (1981) relating post stack time migrated and unmigrated dips to the case of prestack time migration and showed its usefulness in converging fast to an accurate rms velocity model. Both formulae are however restricted to 2D acquisition geometry—we extend the same to 3D case here.

We also work out the 3D migration ellipsoid, and migration aperture in orthogonal directions. The paper is organized as follows. We first derive the formula connecting dips using a planar dipping reflector and calculating the distance between a virtual source (i.e. source reflected in the plane) and a receiver point on a certain acquisition azimuth. Using simple 3D geometry, we then calculate the migration ellipsoid and the migration apertures in two orthogonal directions. We then study the variations of these with different dips and azimuths.

## 3D Relation between dips

We consider a plane $z = x \tan\theta$ with strike in the y-direction i.e. dipping in the x-direction, as shown in Figure 1. We consider a 2D acquisition line shown in green, given by $y = x \tan\phi$ in the plane $z = 0$. In order to calculate finite offset times recorded by this line from the 3D plane reflector, we need to reflect the acquisition line in the plane and calculate distances from the virtual source to the receiver placed at an offset from the source. For a source point $(m, m\tan\phi, 0)$ on the acquisition line, coordinates of the point reflected in the plane $z = x \tan\theta$ are given as

$$\begin{aligned} x_r &= m \cos 2\theta \\ y_r &= m \tan\phi \\ z_r &= m \sin 2\theta \end{aligned} \quad (1)$$

We use $m$ (instead of $x$) to define a point on the acquisition line to distinguish it from a general $x$. Taking distance of the receiver from the source as 2h, coordinates of the receiver are given as

$$(m + 2h\cos\phi, m\tan\phi + 2h\sin\phi, 0) \quad (2)$$

It is now easy to calculate the distance of the virtual source $(x_r, y_r, 0)$ from the receiver. Square of the distance is given as

$$v^2 t^2 = (m\cos 2\theta - m - 2h\cos\phi)^2 + 4h^2 \sin^2\phi + m^2 \sin^2 2\theta, \quad (3)$$

Where $v$ is the assumed constant velocity of the medium and $t$ is the two way time. Simplifying the expression, we get

$$v^2 t^2 = 4(m\sin^2\theta + h\cos\phi)^2 + 4h^2 \sin^2\phi + m^2 \sin^2 2\theta$$
$$= 4m^2 \sin^4\theta + 4h^2 + 8mh\sin^2\theta\cos\phi + 4m^2 \sin^2\theta \cos^2\theta$$
$$= 4(m^2 \sin^2\theta + h^2 + 2hm\sin^2\theta\cos\phi)$$

Implying that

$$t = \frac{2}{v} \sqrt{(m^2 \sin^2\theta + h^2 + 2hm\sin^2\theta\cos\phi)} \qquad (4)$$

In order to find the dip on unmigrated constant offset gather, we need to differentiate time $t$ above, w.r.t. distance $s$ along the acquisition direction $y = m \tan\phi$ i.e.

$$\frac{dt}{ds} = \frac{dt}{dm}\frac{dm}{ds} = \left(\frac{2m\sin^2\theta + 2h\sin^2\theta\cos\phi}{v\sqrt{(m^2\sin^2\theta + h^2 + 2hm\sin^2\theta\cos\phi)}}\right)\frac{dm}{ds} \qquad (5)$$

Using relations $ds^2 = dm^2 + dy^2$ & $dy = dm \tan\phi$, we get $ds = dm \sec\phi$ or

$$\frac{dm}{ds} = \cos\phi \qquad (6)$$

Substituting (6) in (5), we get

$$\frac{dt}{ds} = \left(\frac{2m\sin^2\theta + 2h\sin^2\theta\cos\phi}{v\sqrt{(m^2\sin^2\theta + h^2 + 2hm\sin^2\theta\cos\phi)}}\right)\cos\phi \qquad (7)$$

We now need to establish that formula (7) reduces to the formula for the 2D case (Singh, 2024). We recapitulate briefly below the calculation used in Singh (2024). Refer to figure 3, where we consider a dipping reflector with dip $\theta$. With a source placed at S and a receiver at R, we get reflection from point P' on the reflector, obtained by joining the virtual source S' with R by a straight line. The dotted curve is an ellipse corresponding to a fixed two way travel time $(S'P'R)/v$, where $v$ is the velocity of the constant velocity medium.

Taking $SR = 2h$ and $SS' = d$, we obtain from triangle $SS'R$:

$$(S'P'R)^2 = d^2 + 4h^2 + 4dh\sin\theta \qquad (8)$$

Differentiating (6), we obtain

$$2(S'P'R)\,\delta(S'P'R) = 2d\,\delta d + 4h\,\delta d \sin\theta$$

or

$$(S'P'R)\,\delta(S'P'R) = \delta d\,(d + 2h \sin\theta) \qquad (9)$$

Taking the origin at the point where the dipping reflector and acquisition surface meet (as in the 3D case), we get

$$d = 2x \sin\theta, \qquad (10)$$

so that

$$\frac{\delta(S'P'R)}{\delta x} = \frac{2\sin\theta(d + 2h\sin\theta)}{\sqrt{d^2 + 4h^2 + 4dh\sin\theta}}$$

Or

$$\frac{\delta t}{\delta x} = \frac{2\sin\theta(d + 2h\sin\theta)}{v\sqrt{d^2 + 4h^2 + 4dh\sin\theta}} \qquad (11)$$

Using (10) in (7) (equating $x = m$), it is easy to see that formula (7) for the 3D case reduces to formula (11) for $\phi = 0$, the 2D case, as it should.

## Migration ellipsoid and migration aperture

We now turn our attention to deriving prestack migration aperture for the 3D case for which we need to determine the migration ellipsoid for time given by (4). Taking $x = m$ in (4), multiplying the equation by velocity $v$ and equating it with the sum of distances of a general point $(x, y, z)$ from a source-receiver pair with offset $2h$ on the acquisition line, we get

$$4(m^2 \sin^2 \theta + h^2 + 2hm\sin^2\theta\cos\phi) = \left( \sqrt{(x-m)^2 + (y - m\tan\phi)^2 + z^2} + \sqrt{(x-(m+2h\cos\phi))^2 + (y - (m\tan\phi + 2h\sin\phi))^2 + z^2} \right)^2, \quad (12)$$

where $(m, m\tan\phi)$ represents a point on the acquisition line $y = x\tan\phi$. In order to find migration aperture, we need to find the line connecting points given by reflected virtual source (1) and the receiver (2) for $x = m$ and find its point of intersection with the dipping plane. We modify the equation of plane to $z = x\tan\theta + d'$ so that the receiver is at a non-zero distance $d'$ from the plane when $\theta = 0$ (we use symbol $d'$ to distinguish it from symbol $d$ used earlier in equations 8-11). For modified equation of the plane, $x_r, y_r, z_r$ in (1) is modified to the following

$$x_r = (m(1 - \tan^2\theta) - 2d'\tan\theta) / (1 + \tan^2\theta)$$
$$y_r = m\tan\phi \quad (13)$$
$$z_r = 2(d' + m\tan\theta)\cos^2\theta$$

Equation of line connecting points (13) and (2) is given as

$$\frac{x - (m + 2h\cos\phi)}{2h\cos\phi + m - m(1 - \tan^2\theta)\cos^2\theta + 2d'\tan\theta \cos^2\theta} = \frac{y - (m\tan\phi + 2h\sin\phi)}{2h\sin\phi} = -\frac{z}{2(d' + m\tan\theta)\cos^2\theta} \quad (14)$$

Substituting $z = x\tan\theta + d'$ in the above equation, we get the following equation for the x-coordinate of the intersection point or the point of reflection on the plane:

$$x = \frac{(m + 2h\cos\phi)(2(d' + m\tan\theta)\cos^2\theta) - d'(2m\sin^2\theta + 2h\cos\phi + 2d'\sin\theta\cos\theta)}{2(d' + m\sin\theta)\cos^2\theta + \tan\theta(2h\cos\phi + 2m\sin^2\theta + 2d'\sin\theta\cos\theta)}$$

$$= \frac{\cos2\theta(2d'm + 2dh\cos\phi) + \sin2\theta(m^2 + 2mh\cos\phi - d'^2)}{2d' + m\sin2\theta + 2m\sin^2\theta\tan\theta + 2h\cos\phi\tan\theta} \quad (15)$$

So x-aperture in the 3D case can be found by subtracting the above from CMP's x-ordinate i.e.

$$x_{apert} = m + h\cos\phi - \frac{\cos2\theta(2d'm + 2Dh\cos\phi) + \sin2\theta(m^2 + 2mh\cos\phi - d'^2)}{2d' + m\sin2\theta + 2m\sin^2\theta\tan\theta + 2h\cos\phi\tan\theta}$$

$$= \frac{4d'm\sin^2\theta + 2m^2\sin^2\theta\tan\theta + 2mh\cos\phi\tan\theta(1 + \sin^2\theta + 4d'h\cos\phi\sin^2\theta - hm\cos\phi\sin2\theta + 2h^2\cos^2\phi\tan\theta + d'^2\sin2\theta}{2d' + m\sin2\theta + 2m\sin^2\theta\tan\theta + 2h\cos\phi\tan\theta} \quad (16)$$

Using (15) in (13), y-abscissa corresponding to (15) is found to obey the following equation

$$y = m tan\phi + 2hsin\phi$$

$$- \frac{\left(tan\theta \frac{cos2\theta(2Dm + 2d'hcos\phi) + sin2\theta(m^2 + 2mhcos\phi - d'^2)}{2d' + msin2\theta + 2msin^2\theta tan\theta + 2hcos\phi tan\theta} + d'\right) 2hsin\phi}{(2d'cos^2\theta + 2msin\theta cos\theta)} \quad (17)$$

So that y-aperture is given as

$$y_{apert} = \frac{2hsin\phi}{(2d'cos^2\theta + 2msin\theta cos\theta)}(d' + tan\theta \frac{cos2\theta(2d'm + 2d'hcos\phi) + sin2\theta(m^2 + 2mhcos\phi - d'^2)}{2d' + msin2\theta + 2msin^2\theta tan\theta + 2hcos\phi tan\theta}) \\ - hsin\phi \quad (18)$$

which obviously vanishes for the 2D case i.e. $y_{apert} = 0$ for $\phi = 0$. Note that we have found expressions for apertures in $x$ and $y$ directions i.e. dip and strike directions. Since inline and crossline apertures are the ones that are coded in a migration algorithm, we need to convert the $x$ and $y$ apertures to inline and crossline apertures. This can be done easily by rotation of coordinates of the reflection point given by (15) and (17) to a coordinate system with inline ($y = xtan\phi$) and corresponding crossline as the new axes i.e. $x'$ and $y'$ directions so that

$$\begin{pmatrix} x' \\ y' \end{pmatrix} = \begin{pmatrix} cos\phi & sin\phi \\ -sin\phi & cos\phi \end{pmatrix} \begin{pmatrix} x \\ y \end{pmatrix} \quad (19)$$

Distances of $x'$ and $y'$ from the CMP location which now lies on $x'$ axis would give us the inline and crossline apertures. We have avoided writing out the long and complicated expressions for the same—only results obtained from these will be discussed below.

## Results and Discussion

In Figure 1, we show the dipping plane given by $z = x\, tan\, \theta$ in purple, green line shows the acquisition line in x-y plane at an azimuth $\phi$, blue line shows un-migrated dips for the data recorded along the green line for a constant offset. Red line shows the same for data that what would be recorded by a 2D line along the dip direction i.e. x-axis. In Figure (2), we show that the blue line merges with the red line for $\phi = 0$. In Figure (4), we plot the migration ellipsoid given by (12) at two different positions on the acquisition line and see that the dipping plane is tangent to the ellipsoid at the migrated points. Figure (5) shows the recorded unmigrated data (orange line) along with projection of acquisition line $y = x\, tan\, \phi$ (blue line) on the dipping plane, where the data would lie after migration. In Figure (6), we show the migrated and recorded (unmigrated) data points for a given source and CMP position.

In our formulae, angles $\theta$ and $\phi$ i.e. dip and azimuth appear repeatedly. Using formula (7), ratio of dips in two mutually perpendicular directions $\phi$ and $\phi + \frac{\pi}{2}$ is found to be $tan\phi$ i.e. azimuth can be determined from this ratio. Having determined azimuth, dip of the planar reflector may also be determined using formula (7) from just two mutually perpendicular (L-shaped) zero offset 2D lines intersecting at a point. This L-shaped 2D set-up may be used at multiple points over the surface to get an idea of the 3D subsurface before planning a more elaborate 3D survey for higher resolution and clarity.

Figure 7 shows variation of apertures in $x$ and $y$ directions with offset. Variations of aperture in two directions (dip and strike) for identical parameters with dip and azimuth are shown in Figures (8-10). Figure (11) and (12) show variation of inline and crossline apertures with dip angle for a fixed azimuth and with azimuth for a fixed dip angle respectively.

In order to make the formulae for migration aperture more useful, we need to express apertures given by equations (16-19) in terms of known parameters. We do this by using equation (4) for $vt$ and taking $d' = 0$ in equations (16-19). Taking $d' = 0$ may appear not to cover the most general case, but for a given $vt$, it captures the essential behaviour of inline and crossline apertures i.e. their variations with dip angle, azimuth, and offset, for a given velocity and two way time. We now plot inline aperture as a function of dip angle for a fixed azimuth and fixed $vt$, and inline aperture as a function of azimuth for a fixed dip angle and fixed $vt$ in Figure (13). As expected, aperture increases with dip angle and decreases with azimuth—and is zero along the strike direction i.e. at an azimuth of $\pi/2$. Figure (14) shows variation of inline migration aperture with offset for two different dip angles for an azimuth of $\pi/4$.

## Conclusions

We have extended relation between dips of prestack time migrated and unmigrated constant offset gathers proposed for the 2D case, Singh(2024)) to the 3D case i.e. as a function of both dip and azimuth (formula 7). Using simple considerations, we can determine both dip and azimuth of a planar reflector from data acquired in two perpendicular directions, so that the 3D formula can be used. We have already seen in the earlier paper that this formula allows us faster convergence to the correct rms velocity model—this becomes even more important in the 3D case as considerable time would be saved in velocity picking.

We have also derived exact formulae for inline and crossline migration apertures as a function of dip, azimuth, offset, velocity and two way time. This allows us to use optimum apertures in both directions and avoid artefacts due to use of over or under-aperture, Yilmaz (2001).

## References

Singh Jagmeet, 2024, Relation between Dips of Prestack Time Migrated & Unmigrated Data, The International Meeting for Applied Geoscience and Energy, Houston, Society of Exploration Geophysicists, USA.

Chun, J.H. and Jacewitz, C., 1981, Fundamentals of frequency-domain migration: Geophysics, 46, 717–732.

Oz Yilmaz, 2001, Seismic Data Analysis, Society of Exploration Geophysicists

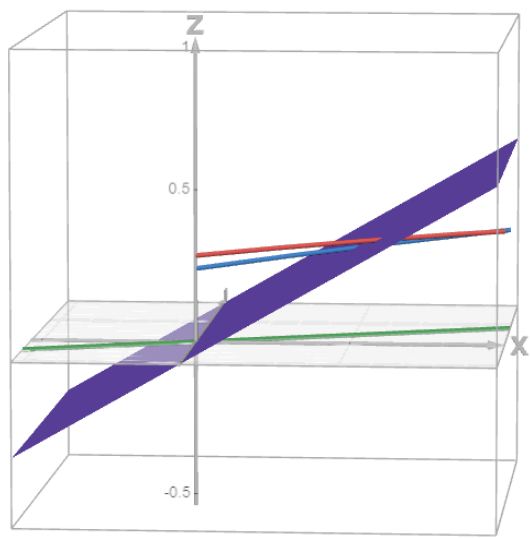

Figure 1: Dipping plane (dipping in x-direction) shown in purple, green line shows an acquisition line in x-y plane at an azimuth, blue line shows un-migrated dips for the data recorded along the green line for a constant offset. Red line shows the same for data that what would be recorded by a 2D line along the dip direction i.e. x-axis.

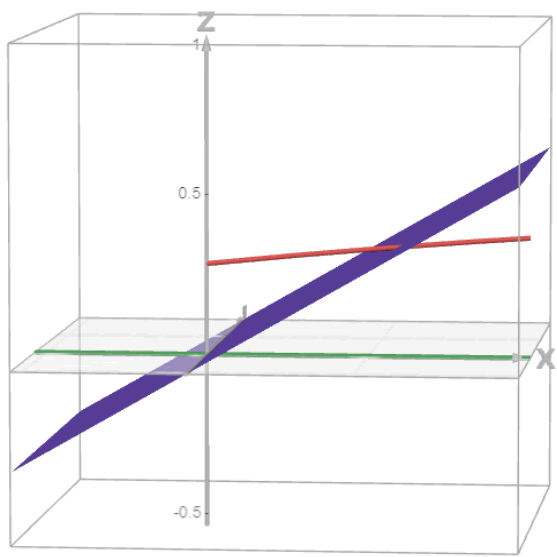

Figure 2: Same as figure 1 except that the green acquisition line is now along the x-axis. Blue and red lines of figure 1 now merge into a single line.

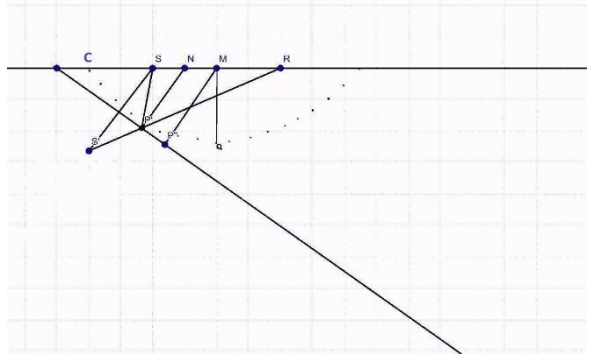

Figure 3: Ray paths and migration ellipse for a dipping reflector for the 2d case.

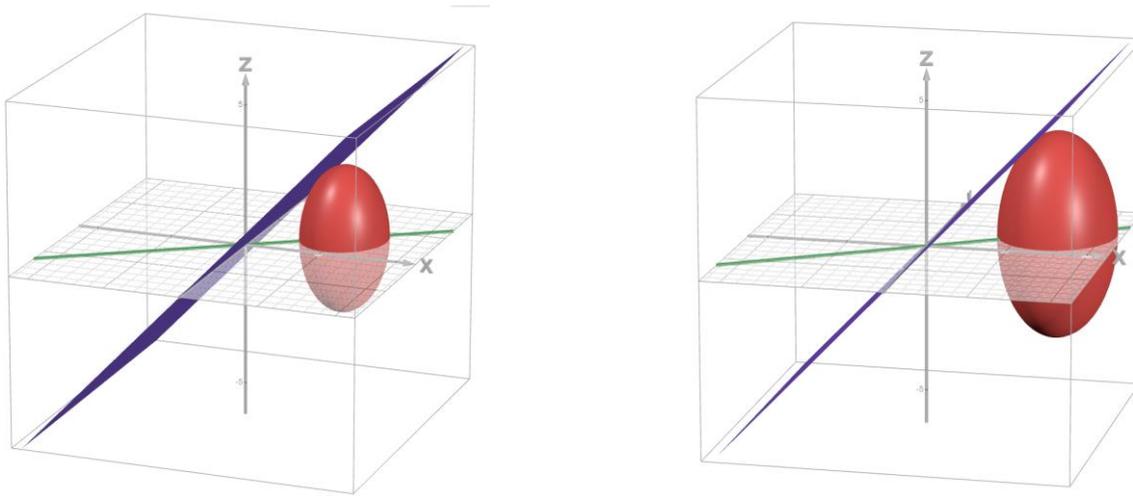

Figure 4: Migration ellipsoid given by equation (10) shown at two different positions on the acquisition line (green).

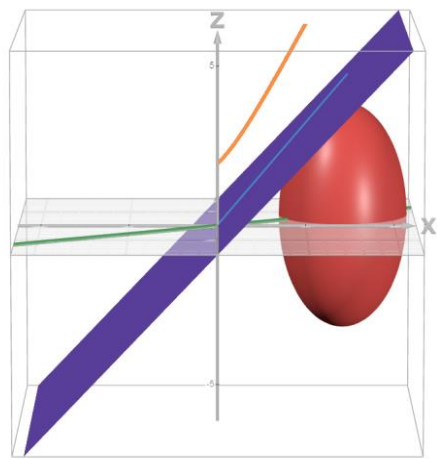

Figure 5: RHS of Figure 4 is shown along with constant offset data (orange curve) that would be recorded by source-receiver pairs on the green acquisition line along with data that would be obtained after 3D migration (light blue line). Data here is shown in terms of two way distances rather than times to show that the migrated data lies on the dipping plane.

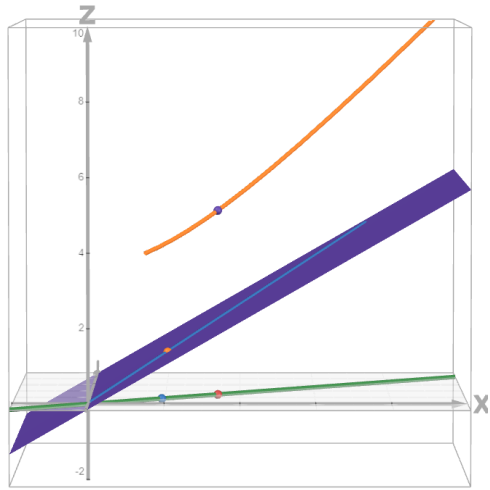

Figure 6: Blue and red dots on the acquisition line (green) show the source position and CMP position respectively. Purple dot on the orange line (recorded data) shows the data point corresponding to the CMP and the orange dot on the dipping plane shows the migrated data point.

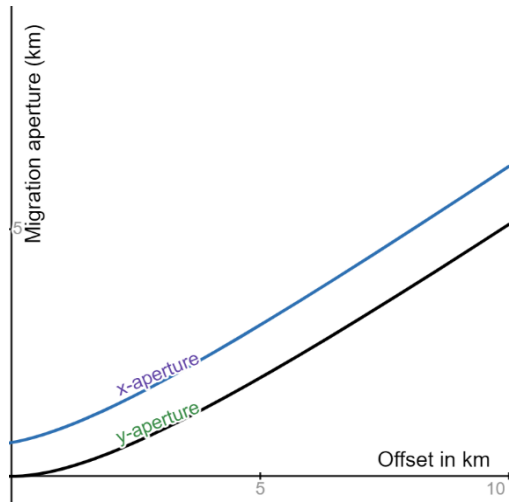

Figure 7: Migration apertures in $x$ and $y$ directions as a function of offset.

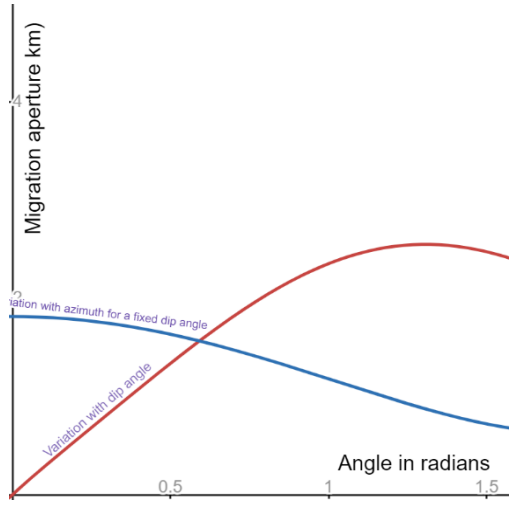

Figure 8: Migration aperture in x-direction: variation with dip angle for a fixed azimuth of $\frac{\pi}{4}$ (red curve) and variation with azimuth for a fixed dip angle 0f 30 degrees for d' = 1 km, h=2 km, m=1.

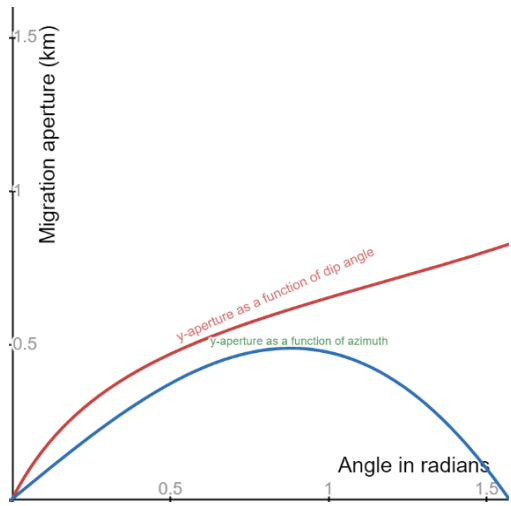

Figure 9: Migration aperture in y-direction: variation with dip angle for a fixed azimuth of $\frac{\pi}{4}$ (red curve) and variation with azimuth for a fixed dip angle 0f 30 degrees for d'= 1 km, h=2 km, m=1.

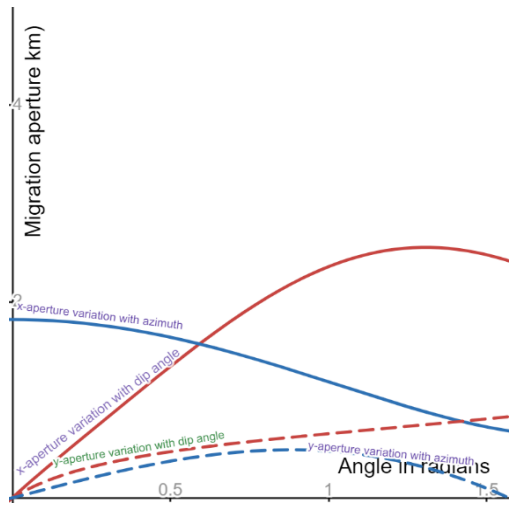

Figure 10: Figures 8 and 9 superposed with dotted and solid curves representing y-aperture and x-aperture respectively.

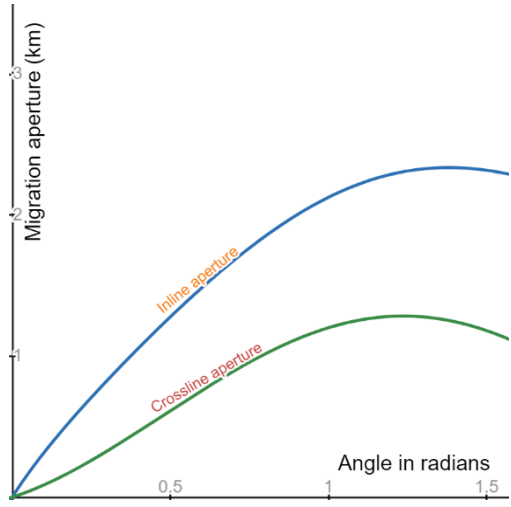

Figure 11: Migration aperture in inline and crossline directions: variation with dip angle for d'= 1 km, h=2 km, m=1 for an azimuth of $\frac{\pi}{4}$

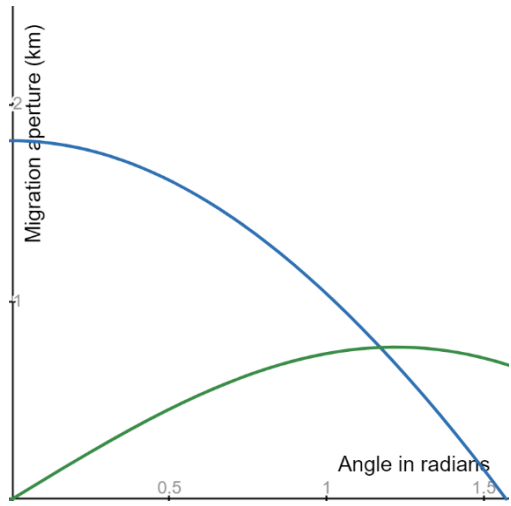

Figure 12: Migration aperture in inline and crossline directions: variation with azimuth for d'= 1km, h=2 km, m=1 for dip angle of $\frac{\pi}{6}$

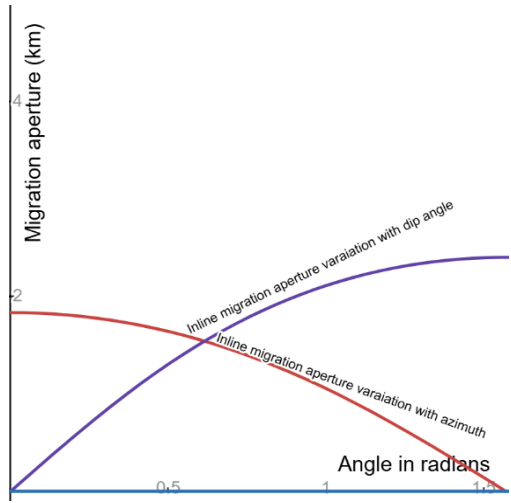

Figure 13: Variation of inline migration aperture with azimuth for dip angle = $\frac{\pi}{6}$ (red curve) and with dip angle for an azimuth of $\frac{\pi}{4}$ (purple curve) for fixed vt =6 km

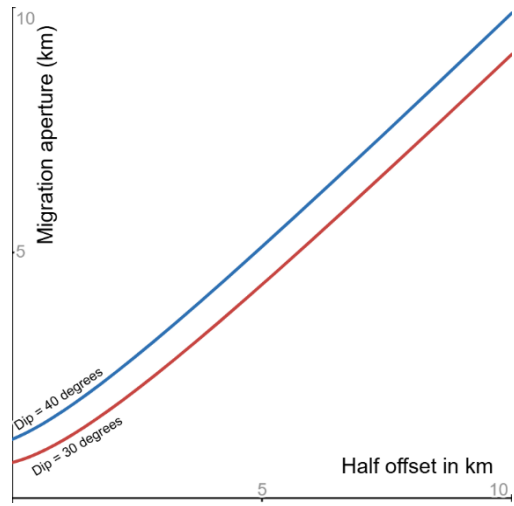

Figure 14: Variation of inline migration aperture with half offset for dips of 40 and 30 degrees for an azimuth of 45 degrees for fixed vt =6 km.